\documentstyle[12pt,graphicx]{article}                
\begin{document}
\baselineskip 18pt
\def\today{\ifcase\month\or
 January\or February\or March\or April\or May\or June\or
 July\or August\or September\or October\or November\or December\fi
 \space\number\day, \number\year}
\def\thebibliography#1{\section*{References\markboth
 {References}{References}}\list
 {[\arabic{enumi}]}{\settowidth\labelwidth{[#1]}
 \leftmargin\labelwidth
 \advance\leftmargin\labelsep
 \usecounter{enumi}}
 \def\newblock{\hskip .11em plus .33em minus .07em}
 \sloppy
 \sfcode`\.=1000\relax}
\let\endthebibliography=\endlist
\def\lsim{\ ^<\llap{$_\sim$}\ }
\def\gsim{\ ^>\llap{$_\sim$}\ }
\def\r2{\sqrt 2}
\def\beq{\begin{equation}}
\def\eeq{\end{equation}}
\def\beqn{\begin{eqnarray}}
\def\eeqn{\end{eqnarray}}
\def\rmuu{\gamma^{\mu}}
\def\rmud{\gamma_{\mu}}
\def\PL{{1-\gamma_5\over 2}}
\def\PR{{1+\gamma_5\over 2}}
\def\sinW2{\sin^2\theta_W}
\def\AEM{\alpha_{EM}}
\def\mul{M_{\tilde{u} L}^2}
\def\mur{M_{\tilde{u} R}^2}
\def\mdl{M_{\tilde{d} L}^2}
\def\mdr{M_{\tilde{d} R}^2}
\def\mz2{M_{z}^2}
\def\c2b{\cos 2\beta}
\def\au{A_u}         
\def\ad{A_d}
\def\cob{\cot \beta}
\def\v#1{v_#1}
\def\tb{\tan\beta}
\def\epem{$e^+e^-$}
\def\KK{$K^0$-$\bar{K^0}$}
\def\wi{\omega_i}
\def\xj{\chi_j}
\def\Wmu{W_\mu}
\def\Wnu{W_\nu}
\def\m#1{{\tilde m}_#1}
\def\mH{m_H}
\def\mw#1{{\tilde m}_{\omega #1}}
\def\mx#1{{\tilde m}_{\chi^{0}_#1}}
\def\mc#1{{\tilde m}_{\chi^{+}_#1}}
\def\mwi{{\tilde m}_{\omega i}}
\def\mxi{{\tilde m}_{\chi^{0}_i}}
\def\mci{{\tilde m}_{\chi^{+}_i}}
\def\mz{M_z}
\def\sw{\sin\theta_W}
\def\cw{\cos\theta_W}
\def\cb{\cos\beta}
\def\sb{\sin\beta}
\def\rwi{r_{\omega i}}
\def\rxj{r_{\chi j}}
\def\rfp{r_f'}
\def\Kik{K_{ik}}
\def\Fq2{F_{2}(q^2)}
\def\f{\({\cal F}\)}
\def\d1{{\f(\tilde c;\tilde s;\tilde W)+ \f(\tilde c;\tilde \mu;\tilde W)}}
\def\tw{\tan\theta_W}
\def\sec2w{sec^2\theta_W}

\begin{titlepage}

\begin{center}
{\large {\bf  Mixing  of the CP Even and the CP Odd Higgs Bosons
and the EDM Constraints}}\\
\vskip 0.5 true cm
\vspace{2cm}
\renewcommand{\thefootnote}
{\fnsymbol{footnote}}
 Tarek Ibrahim  
\vskip 0.5 true cm
\end{center}

\noindent
{\center{ Department of  Physics, Faculty of Science,
University of Alexandria,
 Alexandria, Egypt\footnote{:Permanent address}\\ 
{ Department of Physics, Northeastern University,
Boston, MA 02115-5000, USA} \\
}
\endcenter} 
\vskip 1.0 true cm
\centerline{\bf Abstract}
\medskip
The mixing among the CP even and the CP odd neutral Higgs
bosons of MSSM by  one loop induced effects 
in the presence of  CP phases is investigated using three
different mechanisms to satisfy the EDM constraints,
i.e., a fine tuning of phases, a  heavy sparticle spectrum, and the 
cancellation mechanism. It is shown that if a mixing effect
among the CP even and the CP odd Higgs bosons is observed experimentally,
then it is only the cancellation mechanism that can survive 
under the naturalness constraint. 
\end{titlepage}

\section{Introduction}
Supersymmetric models with softly broken supersymmetry 
contain parameters which are in general complex and 
the phases associated with them are in general O(1).
Such phases induce CP violation and contribute to the electric
dipole moment of the electron and of the neutron and because
the phases are O(1) such contributions are in general very large
and in conflict with the experimental values\cite{commins}:
\beq
|d_e|<4.3\times 10^{-27} e cm, |d_n|<6.3\times 10^{-26} e cm.
\eeq

 There are several solutions
suggested in the literature to avoid this conflict.
One possibility suggested is that the phases could be small
$O(10^{-3})$ while the SUSY spectrum is moderate\cite{ellis}. This 
 possibility, however, constitutes  a 
fine-tuning. Another possiblity is that the EDMs are suppressed kinematically
 because of the heaviness of the sparticle masses that 
 are exchanged in the loops of the EDMs operators\cite{na}. 
 This possibility puts the SUSY particle even beyond the 
 reach of the Large Hadron Collider (LHC) and the
  heavy spectrum itself represents
another fine tuning\cite{fin}. 
However, it was demonstrated that this need not to be the
case and indeed there could be consistency with experiment even with 
large CP phases and a light spectrum
due to an internal cancellation mechanism among the various contributions 
to the EDM\cite{in1}. The above possibility has led to considerable 
further
activity and the effects of large CP phases have been studied in
 different phenomena\cite{in2}.
There are other possibilities suggested
in the literature to overcome the EDM problem which consists of
using a mixture of the previous two 
scenarios\cite{dimop1}. As an example of a mixed
solution is the one with non-universal trilinear coupling
$A_f$  (f=1,2,3). This scenario requires the phase of $\mu$ to be fine
tuned to $O(10^{-2})$ or less,  $phase (A_f) = (0,0,phase (A_3))$ and the gluino mass 
to be heavy. Thus some phases of the theory are fine tuned and others are of
O(1). To suppress the effects of these selected phases one has to push
the corresponding masses up. 

The cancellation mechanism  opens a window of
 physics with large CP phases and a light SUSY spectrum. 
Analyses have been carried out to investigate the  effects of 
CP phases on dark matter,
on $g_{\mu}-2$, on proton lifetime and on other low energy processes. 
One sector of relevance to us here is the neutral Higgs sector. It was pointed out 
in Ref. \cite{pilaftsis} that
the presence of CP phases in the soft SUSY breaking sector
will induce CP  effects in the neutral Higgs sector allowing a mixing
of the CP even and the CP odd states.
Effects of mixings arising from the exchange of the top-stops
and bottom-sbottoms were computed in Ref.\cite{demir}. 
Additional contributions
from the chargino, the W and the charged Higgs exchange 
loops were studied in
Ref. \cite{uswe}. 

In this paper we use the CP properties of the neutral Higgs bosons as
an experimental probe to compare  the three main solutions to 
EDM problem. The solutions of mixed type will not be considered here in details 
since their behaviour could be inferred from the main scenarios. 
We show that among the three scenarios, the
cancellation mechanism
has somewhat of a unique  position in that it is the favored
 solution 
 to the 
 EDM problem if the MSSM Higgs bosons are discovered and  are found
 to be admixtures of CP even and CP odd states.

To evaluate the radiative corrections to the Higgs boson masses 
and mixings we use the effective potential approximation
\beq
V=V_0+\Delta V\nonumber\\
\eeq
where $V_0$ is the tree-level potential and $\Delta V$
is the one loop Coleman-Weinberg correction
\beq
\Delta V=\frac{1}{64\pi^2}
 Str(M^4(H_1,H_2)(log\frac{M^2(H_1,H_2)}{Q^2}-\frac{3}{2})
\eeq
with $Str=\sum_i C_i (2J_i+1)(-1)^{2J_i}$
where the sum runs over all particles with spin $J_i$ 
and $C_i(2J_i+1)$ counts the degrees of the particle i,
 Q is the running scale and $H_{1,2}$ are the SU(2) Higgs doublets
with non vanishing vacuum expectation $v_1$ and $v_2$: 
\beqn
(H_1)= \left(\matrix{H_1^0\cr
 H_1^-}\right)
 =\frac{1}{\sqrt 2} 
\left(\matrix{v_1+\phi_1+i\psi_1\cr
             H_1^-}\right)\nonumber\\
(H_2)= \left(\matrix{H_2^+\cr
             H_2^0}\right)
=\frac{e^{i\theta_H}}{\sqrt 2} \left(\matrix{H_2^+ \cr
             v_2+\phi_2+i\psi_2}\right)
\eeqn

We consider here the contributions from the top-stop, 
and the bottom-sbottom and from the W-charged Higgs-chargino sector
exchange. The mass squared matrix of 
the neutral Higgs bosons is defined by
\beq
M_{ab}^2=(\frac{\partial^2 V}{\partial \Phi_a\partial\Phi_b})_0
\eeq
where $\Phi_a$ (a=1-4) are defined by
\beq
\{\Phi_a\}= \{\phi_1,\phi_2, \psi_1,\psi_2\}
\eeq
and the subscript 0  means that we set 
$\phi_1=\phi_2=\psi_1=\psi_2=0$.
  We introduce a 
 new basis $\{ \phi_1,\phi_2,\psi_{1D}, \psi_{2D}\}$
where $\psi_{1D},\psi_{2D}$ are defined by 
\beqn
\psi_{1D}=\sin\beta \psi_1+ \cos\beta \psi_2\nonumber\\
\psi_{2D}=-\cos\beta \psi_1+\sin\beta \psi_2
\eeqn
and where $\tan \beta=\frac{v_2}{v_1}$.
In this basis the field $\psi_{2D}$ decouples from the other three
fields and is a massless state (Goldstone field).
 The
  Higgs $(mass)^2$ matrix $M^2_{Higgs}$ of the remaining three 
fields is given
  by
\beq
M^2_{Higgs}=
\left(\matrix{M_Z^2c_{\beta}^2+M_A^2s_{\beta}^2+\Delta_{11} &
-(M_Z^2+M_A^2)s_{\beta}c_{\beta}+\Delta_{12} &\Delta_{13}\cr
-(M_Z^2+M_A^2)s_{\beta}c_{\beta}+\Delta_{12} &
M_Z^2s_{\beta}^2+M_A^2c_{\beta}^2+\Delta_{22} & \Delta_{23} \cr
\Delta_{13} & \Delta_{23} &(M_A^2+\Delta_{33})}\right)
\eeq
where $(c_{\beta}, s_{\beta})=(\cos\beta, \sin\beta)$.
The detailed structure of the elements in the above matrix  is given in 
Ref. \cite{uswe}.

We note that the basis fields $\{ \phi_1,\phi_2,\psi_{1D}\}$ 
of matrix (8) 
are the real parts of the neutral Higgs fields and a linear
combination of their imaginary parts
 as displayed in Eq. (4) and so these basis fields are  pure
CP states. Thus $\phi_{1,2}$ are 
CP even (scalars) and $\psi_{1D}$ is CP odd (a pseudoscalar).
What we are interested here is the mixing between the CP
 even and the CP odd Higgs states in the eigen vectors of Eq. (8)
 and this mixing is governed by the off diagonal elements $\Delta_{12}$
and $\Delta_{23}$. These can be written in the form
\beq
\Delta_{12,23}=A_{12,23}\sin\gamma_t
+B_{12,23}\sin\gamma_b
+C_{12,23}\sin\gamma_2
\eeq
where $A$, $B$ and $C$ are  functions of
the masses, of couplings, and of other SUSY parameters such as
$\tan\beta$, etc.,
and  $\gamma_{t,b,2}$ are linear combination
of
the CP  phases and are given by
\beq
\gamma_t=\alpha_{A_t} + \theta_{\mu}, ~~
\gamma_b=\alpha_{A_b} + \theta_{\mu},~~ 
\gamma_2=\xi_2 + \theta_{\mu},
\eeq
where $\alpha_{A_f}$ is the phase of $A_f$, $\theta_{\mu}$ is the
phase of $\mu$ parameter in the superpotential and $\xi_2$ is the
phase of the gaugino mass $m_2$.
In the limit of vanishing CP phases the matrix
elements $\Delta_{12}$ and $\Delta_{23}$
vanish  and thus the Higgs $mass^2$ matrix factors into
a $2\times 2$ CP even Higgs matrix times a CP odd element.
From the structure of the matrix we see that the physical Higgs fields
(the mass eigen states) have mixings between 
their CP even and their CP odd components and these mixings
are induced purely by the existence of CP  phases.
Diagonalizing the $M^2_{Higgs}$ matrix
\beq
RM^2R^T=diag(m^2_{H_1},m^2_{H_2},m^2_{H_3}
)
\eeq
one can calculate the percentage of the CP even components $\phi_{1,2}$
and the CP odd component $\psi_{1D}$ of the three physical fields 
$H_{1,2,3}$. We order the eigen values so that in
the limit of no mixing between the CP even and the CP odd states one has
$(m_{H_1}, m_{H_2}, m_{H_3})$ $\rightarrow$
$(m_H, m_h, m_A)$ 
where 
$m_H$ is the mass of the heavy CP even state,
$m_h$ the mass of the light CP even and $m_A$ is the mass
of the CP odd Higgs in MSSM when all CP  phases are set to zero.
The similarity of the dependence on $\sin(SUSY phases)$ of the
mixing matrix elements $\Delta_{12,23}$ and of the EDMs\cite{na,in1,in2}
 is striking and
indicates that there is a strong correlation between these two physical
phenomena i.e. fermions of the theory possess EDMs and the 
neutral Higgs sector of the theory consists of non-pure CP states.

\section{EDM and Higgs mixings analysis}

We begin with a brief discussion of the EDMs of the electron and of 
the neutron. For the case of the electron we have that 
the electric dipole
moment operator has only two components, i.e., the loop contribution from
the chargino and from the neutralino . For the neutron EDM, we have three
operators, the chromoelectric dipole moment , the electric dipole
moment  and the purely gluonic dimension six operator. The
chromoelectric and the electric operators each 
have three components arising from the chargino, the neutralino and 
the gluino exchange loops.
In addition to the above, there  are two loop
graphs\cite{twol} which contribute to the EDMS and have been 
included in the present analysis. 
However, it turns out that in essentially all regions 
of the parameter space investigated, the effects of the
two loop contributions is very small.

For the general analysis in MSSM the number of parameters that are involved
in the EDM and in the Higgs mixings is rather large. 
For the purpose of
numerical study we confine ourselves to a constrained set consisting of:
 $m_0, m_{\frac{1}{2}}$, $m_A$, $|A_0|$, $\tan\beta$,
$\theta_{\mu}$, $\alpha_{A_0}$, $\xi_1$, $\xi_2$ and $\xi_3$.
(For definition of the parameters see Refs.[6] and [11]).
Using this parameter set we generate the sparticle masses 
at low energy starting
at the GUT scale and evolving them down to the
electroweak scale using renormalization group analysis. The 
parameter $\mu$ in the superpotential is to be evaluated 
using the constraint of radiative breaking of the  electroweak
symmetry. 

In addition to the masses of sparticles and
 CP  phases that affect the magnitude of the 
EDMs and the Higgs mixings,  $\tan\beta$  has a large
effect on the EDMs and also plays a crucial role in 
Higgs mixings as will be seen later. 
In Fig.1 and Fig.2 we show  
the $\tan\beta$ dependence of the different components of the
 electron
 and of the neutron EDMs respectively. 
  The figures show that
the magnitudes of the electron and of the neutron EDM
 generally increase
as $\tan\beta$ increases. For the case of the electron in Fig. 1
 this behavior can be easily 
understood  since the chargino component has a factor
of $\frac{1}{\cos\beta}$ which grows as $\tan\beta$ increases and also there is another weaker dependence on $\tan\beta$ 
coming from the diagonalizing matrices of the chargino mass
matrix but the net effect of both factors is
to increase the chargino contribution with $\tan\beta$ as shown by 
the solid curve 
of Fig. 1.
The large terms in the neutralino contribution are proportional to $\tan\beta$ 
 coming from the elements of the matrix that diagonalizes the
selectron $mass^2$ matrix. In comparison the contribution which
comes from the elements of the matrix that diagonalizes the
neutralino mass matrix produces only a weak dependence on $\tan\beta$. The neutralino contribution dependence
on $\tan\beta$ is shown by the dashed curve of Fig. 1. 
Thus together the chargino and the neutralino contributions 
 exhibit a dependence on $\tan\beta$ so that their total contribution
increases with increasing $\tan\beta$.
For the neutron case, we have contributions from the up quark and from
the down quark. The up and down quarks have different $\tan\beta$
dependences. In the electric and the chromoelectric dipole moment operators
the up quark contribution decreases as $\tan\beta$ increases while
the down quark contribution increases. However, the down quark
contribution dominates and thus the contribution of these two operators
is an increasing function of $\tan\beta$ as shown by 
the solid and dotted with squars curves  of Fig. 2. The gluonic operator has top-stop
vertex
in its two loop diagrams and thus has a term of $\cot\beta$ that decreases
as $\tan\beta$ increases but it has also a bottom-sbottom
vertex that contributes a linear dependence on $\tan\beta$.
Thus $d^G$ has a more complicated dependence on $\tan\beta$ as
shown by the dashed curve of Fig. 2, where the sbottom contribution
 is larger in that area of parameter space.
 However, unless 
there are large cancellations 
between the electric and the chromoelectric operators the 
contribution from
the gluonic operator is relatively small. As a result the 
neutron EDM is also
an increasing function of $\tan\beta$.

 We discuss now the Higgs mixing. We use the 
matrix elements $R_{ij}$ of the diagonalizing matrix $R$ in Eq. 
(11) to calculate the percentage of the CP even and the CP odd components
of the physical Higgs fields in the neutral sector.
In most of the parameter space investigated in Ref. \cite{demir} and \cite{uswe},
$H_2$, which limits to the CP even lightest Higgs as the
CP  phases vanish, develops almost no CP odd component and
thus remains a pure CP even state. This result can be understood from
the large difference in the magnitude between the two mixing
terms $\Delta_{13}$ and $\Delta_{23}$. Thus MSSM with phases has
one pure scalar Higgs field and the other states, i.e., $H_1$ and
$H_3$, can have mixings. This mixing in $H_1$ and $H_3$  
 depends on the size of the 
CP phases and on $\tan\beta$, as demonstrated
in Refs. \cite{demir} and \cite{uswe}. The effect of  
$\tan\beta$ on the mixing is dramatic. 
This is  shown in 
  Fig. 3, where we have plotted the CP odd and the even components of $H_1$
as functions of $\tan\beta$. It is clear that even with large
CP phases the CP odd component of $H_1$ is very small for
 $\tan\beta<15$. However, this component can get
  very large as $\tan\beta$ gets large and it could
become as much as $50\%$ when $\tan\beta=30$. The
$\tan\beta$  dependence
of  the CP even-odd components of $H_1$ and $H_3$  
 can be understood
from the matrix (8).
Thus as $\tan\beta$ gets large, $\cos\beta$ becomes very small and
$\sin\beta$  approaches $1$ and the two heavy eigen
masses $m^2_{H_1}$ and $m^2_{H_3}$ become approximately equal.
This degeneracy is accompanied by large $\phi_1$
and $\psi_{1D}$ components in the linear combination structure of 
the two eigen states $H_1$ and $H_3$. 
The analysis of mixing of CP even-odd mixing in $H_3$ is
exactly the same as for $H_1$  except that the components $\phi_1$ and
$\psi_{1D}$  are interchanged in the linear
combinations that appear in $H_1$ and $H_3$.

We explore now the different scenarios for solving the EDM problem
and investigate the behavior of the Higgs mixings
 in each. The relevant variables here are the sparticle masses,
 $\alpha_{A_0}$, $\theta_{\mu}$ and $\xi_i$ with $i=1,2,3$ and
$\tan\beta$.
\begin{table}[h]
\begin{center}
\caption{{\bf EDMs and CP structure of $H_1$ for$|A_0|=2$, $m_A=500$}}
\begin{tabular}{|l|l|l|}
\hline
\hline
$(case)\tan\beta$,$m_0$, $m_{\frac{1}{2}}$,\\
 $\xi_1$, $\xi_2$,$\xi_3$,$\theta_{\mu}$, $\alpha_{A_0}$, & $d_e$, $d_n$ & $CP-even$, $CP-odd$\\
\hline
\hline
$(1)2,500,400$,\\
all phases $= 10^{-2}$ & $-7.4\times 10^{-28}$, $-3.1\times 10^{-26}$ & $99.999\%$, $.001\%$ \\
\hline
\hline
$(2)30,500,400$,\\
all phases $= 10^{-3}$ & $-1.3\times 10^{-27}$, $-4.0\times 10^{-26}$ & $99.999\%$, $.001\%$ \\
\hline
\hline
$(3)5,4000,1200$,\\
all phases $= 10^{-1}$ & $-6.2\times 10^{-28}$, $-3.8\times 10^{-26}$ & $99.996\%$, $.004\%$ \\
\hline
\hline
$(4)5,5000,2240$,\\
$-2.,-.3,.5,-.4,.4$ & $1.1\times 10^{-27}$, $2.7\times 10^{-26}$ & $99.86\%$, $.14\%$ \\
\hline
\hline
$(5)30,5000,2240$,\\
all phases $= 10^{-2}$ & $-2.2\times 10^{-28}$, $-9.5\times 10^{-27}$ & $99.996\%$, $.004\%$ \\
\hline
\hline
$(6)30,9500,3600$,\\
$-2.,-.3,.5,-.4,.4$ & $2.2\times 10^{-27}$, $6.2\times 10^{-26}$ &
$ 93.7\%$, $6.3\%$ \\
\hline
\hline
\end{tabular}
\end{center}
\end{table}
\noindent

These three scenarios to solve the EDM problem
 are the fine tuning of phases, the heavy sparticle spectrum, and the
 cancellation mechanism.
 The first scenario for the EDM suppression assumes that the
sparticles could have moderate masses while the CP phases
are very small. An analysis of this scenario is given in cases 1 and 2 
of Table. 1.  Here the CP even-odd mixing in $H_1$ is small
for both small $\tan\beta$ (case 1) and large $\tan\beta$ (case 2)
which is expected since the CP phases are small.

In the second scenario with heavy sparticle spectrum 
the CP phases and $\tan\beta$ could be either small or large
giving rise to four possibilities, i.e.,(1) CP phases small-$\tan\beta$ small,
(2) CP phases large-$\tan\beta$ small, (3) CP phases small- $\tan\beta$ large,
and (4) CP phases large- $\tan\beta$ large.
 We display these four possibilities as the cases
3, 4, 5 and 6 of Table.1. For case 3 
where phases are small and for case 4 where phases are large, we
see that EDMs constraints are satisfied while there is no mixing in the
structure of the $H_1$ field. This field is almost a pure scalar which
is due to the smalleness of $\tan\beta$. It is clear here that 
$\tan\beta$ is a 
crucial parameter in the  Higgs mixing. For case 5 the mixing is
also negligible although $\tan\beta$ is large and that is because
the other important parameters here for mixing to occur i.e 
the CP phases, are small. So from the analysis of the
 three cases 3, 4 and 5 of Table.1, we find that one must have both 
$\tan\beta$ and
CP phases large for the mixing in the Higgs sector to be observable.
For case 6 
where we choose both the phases and $\tan\beta$ to be large 
we find that the $H_1$ state is a mixed state.
We notice that the  sparticle masses involved here are
very heavy and as an exmple the corresponding gluino mass for case
6 is 10.2 TeV.
 The values of $m_0$ and $m_{1/2}$ are the lower bound of these masses for EDMs to be suppressed below the experimental limit.
The CP odd percentage of $H_1$ is $6\%$ and to have more
mixing, we have to make $\tan\beta$ larger and correspondingly also push
the masses up to accommodate the EDM constraints.
Thus for a sizable mixing to occur in the second scenario
 we must have CP phases and as well as $\tan\beta$ to be large 
 which pushes the SUSY
spectrum in the several TeV region to satisfy the EDM constraints.

One may now consider scenarios which are mixtures of the above 
two types and investigate the CP composition of the neutral Higgs in
these. We will consider here two  examples, i.e., the  
flavour-off-diagonal scenario and the focus point scenario. 
In the flavour-off-diagonal scenarios\cite{f-off-d}, 
we have a zero phase for the $\mu$ parameter and for the gaugino masses,
large phases $\alpha_{A_t}$ and $\alpha_{A_b}$ in the third family,
and small phases for $A_u$ and $A_d$ generated by RG effects because
of coupling with the third family.
In this model, the operator $d^G$ makes the dominant contribution 
in the neutron edm and its magnitude is governed by the phases  
$\alpha_{A_t}$ and $\alpha_{A_b}$.  
Between the two  phases
the effect of $\alpha_{A_t}$ on the Higgs mixing 
is larger than the effect of $\alpha_{A_b}$  due to the fact that 
the chargino and stop exchange loop contributions are 
larger than the  sbottom exchange contributions.
Nonetheless it is important  to include both $\alpha_{A_t}$ 
$\alpha_{A_b}$ for a complete analysis of the EDM constraints. 
Having included the sbottom and stop contributions to $d^G$
of the neutron, one finds that for large $\alpha_{A_b}$ and
$\alpha_{A_t}$ phases, $\tan\beta$ should not get large since
sbottom contribution in this case dominates and one needs to 
push gluino mass up to higher values. So if we have a
reasonable range of gluino masses, we find that 
this model gives us a case
for moderate $\tan\beta$ and large phases which is a more
complicated version of case 4 in Table. 1, 
and in this type of models, the neutral Higgs bosons are almost pure CP states.
 For the case where the gluino mass is very large $\tan\beta$ could get 
 larger and in this circumstance the neutral Higgs bosons start to 
 have a mixed CP structure.
In focus point scenario models one can have sfermions masses
of the first two generations in the multi-TeV region without 
affecting the fine tuning parameter of electroweak symmetry 
breaking. Here we can have moderate phases consistent with 
EDM constraints. However for $\tan\beta>10$ \cite{dimop1} 
one cannot satisfy the experimental limits without assuming
 unnaturally small phases angles. 
So here also we have moderate $\tan\beta$ and
large phases. Thus  in this scenario also,
 the neutral Higgs bosons would be almost pure CP states.

Finally, we turn to the third possibility where one has
moderate size masses and large phases which is normally
excluded by the EDM constraints unless we have
simultaneous cancellations among the different components of the electron
and of the neutron EDMs. This region of the parameter space belongs 
to the the mechanism of EDM suppression by cancellations
which we discuss now in detail.
The algorithm to find a simultaneous cancellation of the electron
and of the neutron EDM is straightforward. For the case of the electron
one finds that the chargino component is independent of $\xi_1$ and
$d_e$ as a whole is independent of $\xi_3$. Thus for a given
set of parameters except $\xi_1$ we start varying $\xi_1$ until we
reach the cancellation for $d_e$ since only one of its components
(the neutralino) is affected by this parameter. Once the electric
dipole moment constraint $d_e$ is satisfied we vary $\xi_3$ which
affects only $d_n$ keeping all other parameters fixed. By using this
simple technique one can generate any number of simultaneous cancellations.
It is important to note that the cancellation mechanism also
requires an adjustment of phases. The three important phases
in EDM analysis $\xi_2$, $\xi_3$ and $\theta_{\mu}$ are
to be significantly adjusted for the EDMs to  simultaneously obey
the constraints of the current experimental limits. 
 One hopes that, eventually when
one learns how supersymmetry breaks in string theory, such breaking 
will determine the phases and select the right
mechanism for EDM suppression in SUSY theory.
In the analysis of the cancellation mechanism  given 
in ref.\cite{scal}, EDMs were
shown to obey a simple approximate scaling under the transformation
$m_0\rightarrow \lambda m_0$, $m_{1/2}\rightarrow \lambda m_{1/2}$
in the region where $\mu$ itself obeys the same scaling i.e,
$\mu\rightarrow \lambda \mu$.
In this scaling region, the chargino and the neutralino
contributions for the electric dipole moment of the
electron behave as $EDM \rightarrow \frac{1}{\lambda^2} EDM$
and thus $d_e$ has the same scaling property.
For the neutron case we have that both the electric dipole
moment and the chromoelectric dipole moment operators have
the same scaling  i.e, $d^{E,C}\rightarrow \frac{1}{\lambda^2}
d^{E,C}$ while the gluonic operator has the scaling 
property $d^G \rightarrow \frac{1}{\lambda^4} d^G$.
Thus the scaling property of $d_n$ is more complicated.
However as $\lambda$ gets large the contribution of
$d^G$ falls faster than $d^E$ and $d^C$ and in this
case $d_n\rightarrow \frac{1}{\lambda^2} d_n$
.
 In the scaling region knowledge of a single
point in the MSSM parameter space where cancellation in the EDMs occurs
allows one to generate a trajectory in the $m_0-m_{1/2} $ plane where
the cancellation mechanism holds and the EDMs are small.
The cancellation mechanism can accommodate the EDM constraints with large
CP phases, large or small $\tan\beta$, and a light sparticle spectrum
which could be accessible at colliders.
In Table.2 we display cases 1,2 and 3 where there is a large
mixing. We note that for large $\tan\beta$ 
and large CP  phases the individual components of
the EDMs are large and the suppression here is due to cancellations.
The mixing in $H_1$  reaches $44\%$ in case 3 which is rather large. 
In Fig. 4, we generate the points
of the curves out of points 1, 2 and 3 of Table. 2 by scaling where the
EDMs constraints are satisfied. We have plotted the CP components of
$H_1$ as functions of the scale $\lambda$. 
 We note that the CP even-odd composition of $H_1$ 
is almost independent of scaling. This can be understood 
from the fact that the CP 
 mixing is determined by the dimensionless matrix components $R_{ij}$
which are almost scale independent.

 Because of the $\tan\beta$ dependence of the EDMs  explained above in Figs. 1 and 2,
one finds that in
regions where the gluonic operator
 contribution is small there could be another
scaling. This scaling consists in  multiplying $\tan\beta$ 
by a factor $S$ which generates
another point of cancellation out of a given one. One can also 
generate another
point of cancellation even with gluonic contribution  in
the same order as the other operators if we choose a
value of $S$ close to $1$. 
The important point here is 
that the CP even and the CP odd components  are sensitive functions of
$\tan\beta$. Point 4 is generated from point 1 by multiplying
$\tan\beta$ with $S=\frac{4}{3}$. We note that the  CP odd
component is almost doubled  
while the magnitude of EDMs is still in the
cancellation region.

	We turn now to investigate the implications if MSSM Higgs
	bosons are discovered and are observed to possess
	a sizable CP even-odd mixing. In this case one must have
	both the CP phases large and a large $\tan\beta$.
	Now in this circumstance one can satisfy the EDM constraints
	either by having a sparticle spectrum  in the several TeV 
	region, which  is
 discourgating from the point of view of their observation at colliders,
or the cancellation mechanism which is favorable for the observation
of sparticles at colliders. 
In the latter  case the two Higgs bosons that have maximal mixing 
 have degenerate masses for large $\tan\beta$ as discussed 
 above. We wish to point out, however, that the cancellation 
 mechanism remains a valid scenario even if
the MSSM Higgs bosons are found and observed to be 
 pure CP states as can be seen from case(5) in Table 2. 
 In this region and
because of strong dependence of mixings on phases some large phases
could bring mixings to its minimum value.
       
\noindent
\begin{table}[h]
\begin{center}
\caption{{\bf EDMs satisfied and CP structure of $H_1$}}
\begin{tabular}{|l|l|l|}
\hline
\hline
$m_A$,$\tan\beta$,$m_0$, $m_{\frac{1}{2}}$,$|A_0|$,\\
 $\xi_1$, $\xi_2$, $\xi_3$,$\theta_{\mu}$, $\alpha_{A_0}$, & $d_e$, $d_n$ & $CP-even$, $CP-odd$\\
\hline
\hline
$(1)300,30,500,400,2$,\\
$-2,-.24,-.43,.3,-.4$ & $1.1\times 10^{-27}$, $2.4\times 10^{-26}$ & $88.4\%$, $11.6\%$ \\
\hline
\hline
$(2)400,40,400,490,2$,\\
$-1.2,.57,-.09,-.4,.6$ & $2.1\times 10^{-27}$, $-2.1\times 10^{-26}$ & $68.8\%$, $31.2\%$ \\
\hline
\hline
$(3)300,45,600,600,2$,\\
$.1,-.45,-.28,.4,.5$ & $-3.2\times 10^{-27}$, $2.7\times 10^{-26}$ & $56.4\%$, $43.6\%$ \\
\hline
\hline
$(4)300,40,500,400,2$,\\
$-2,-.24,-.43,.3,-.4$ & $2.3\times 10^{-27}$, $-1.8\times 10^{-26}$ & $79.3\%$, $20.7\%$ \\
\hline
\hline
$(5)300,35,650,400,1$,\\
$-1.5,1.2,1.22,1.95,.7$ & $2.9\times 10^{-27}$, $-1.8\times 10^{-26}$ & $99.94\%$, $.06\%$ \\
\hline
\hline
\end{tabular}
\end{center}
\end{table}
\noindent

In the analysis presented above we did not include the next to leading order 
corrections\cite{new1} in the Higgs mixings.
 However, inclusion of these refinements would not change
  very much the CP structure of the neutral Higgs bosons or the
   conclusions of this paper
\section{Conclusions}

In this paper we have investigated the  effects
of CP phases, of $\tan\beta$ and of the
 sparticles masses on the electron and on the neutron EDMs and 
 on the Higgs mixings.
We have studied three main solutions to the EDM problem and
their implications on the CP properties of the neutral Higgs bosons.
It is found that unless we push the sparticle masses up to
tens of TeV the only way out of the EDM problem is the cancellation
mechanism if the neutral MSSM Higgs bosons are discovered with any
observable CP mixings and one respects the naturalness constraints
that the sparticle masses not lie in the several TeV range.

\noindent
{\bf Acknowledgments}\\ 
I wish to acknowledge useful discussions with Prof. Pran Nath.
The encouraging support of the Physics Department at Alexandria University 
is also acknowledged.

{\bf Figure Captions}\\
Fig.1: Plot of $log_{10}|d_e|$ components as
functions of $\tan\beta$; the solid curve  represents chargino contribution and the dashed curve  is the
neutralino:   
the input parameters are $m_A=300$, $m_0=500$, $m_{\frac{1}{2}}=400$, $|A_0|=1.0$, $\alpha_0=0.5$,
$\xi_1=0.3$, $\xi_2=0.2$, $\xi_3=0.1$ and $\theta_{\mu}=.4$,
where all masses are in GeV and all
angles are in rad.\\

\noindent
Fig. 2: Plot of $log_{10}|d_n|$ components as functions of
$\tan\beta$; the solid curve  represents electric dipole operator, the  
curve  with squars is the chromoelectric operator and the dashed curve  is the
purely gluonic operator contribution  for the same input parameters of Fig. 1.\\

\noindent
Fig.3: Plot of the CP even component $\phi_1$ of $H_1$ (solid 
curve)  and the CP odd component $\psi_{1D}$ of $H_1$ (dashed curve)
including the stop, the sbottom and the chargino sector contributions 
  as a function of $\tan\beta$  for the same inputs as in Fig.1. \\

\noindent
Fig.4: . Plot of the CP even component $\phi_1$ of $H_1$ (dashed
curves)  and the CP odd component $\psi_{1D}$ of $H_1$ (solid curves)
including the stop, the sbottom and the chargino sector contributions 
  as a function of the scale $\lambda$. The inputs for the
far left points of all cuves are those of
points $1,2,3$ of Table. 2 and then scale both $m_0$ and $m_{1/2}$ as one
goes to the right.\\

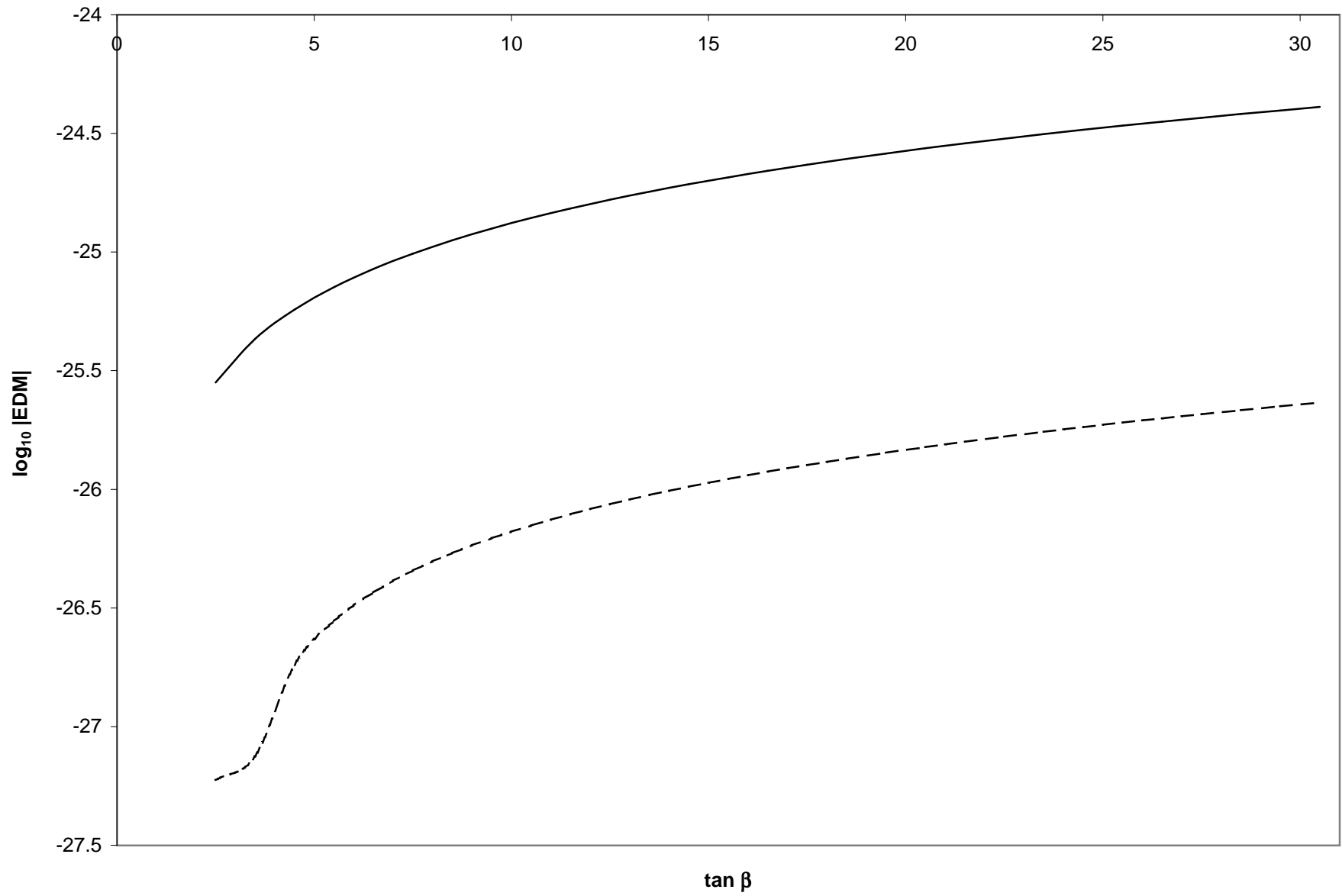
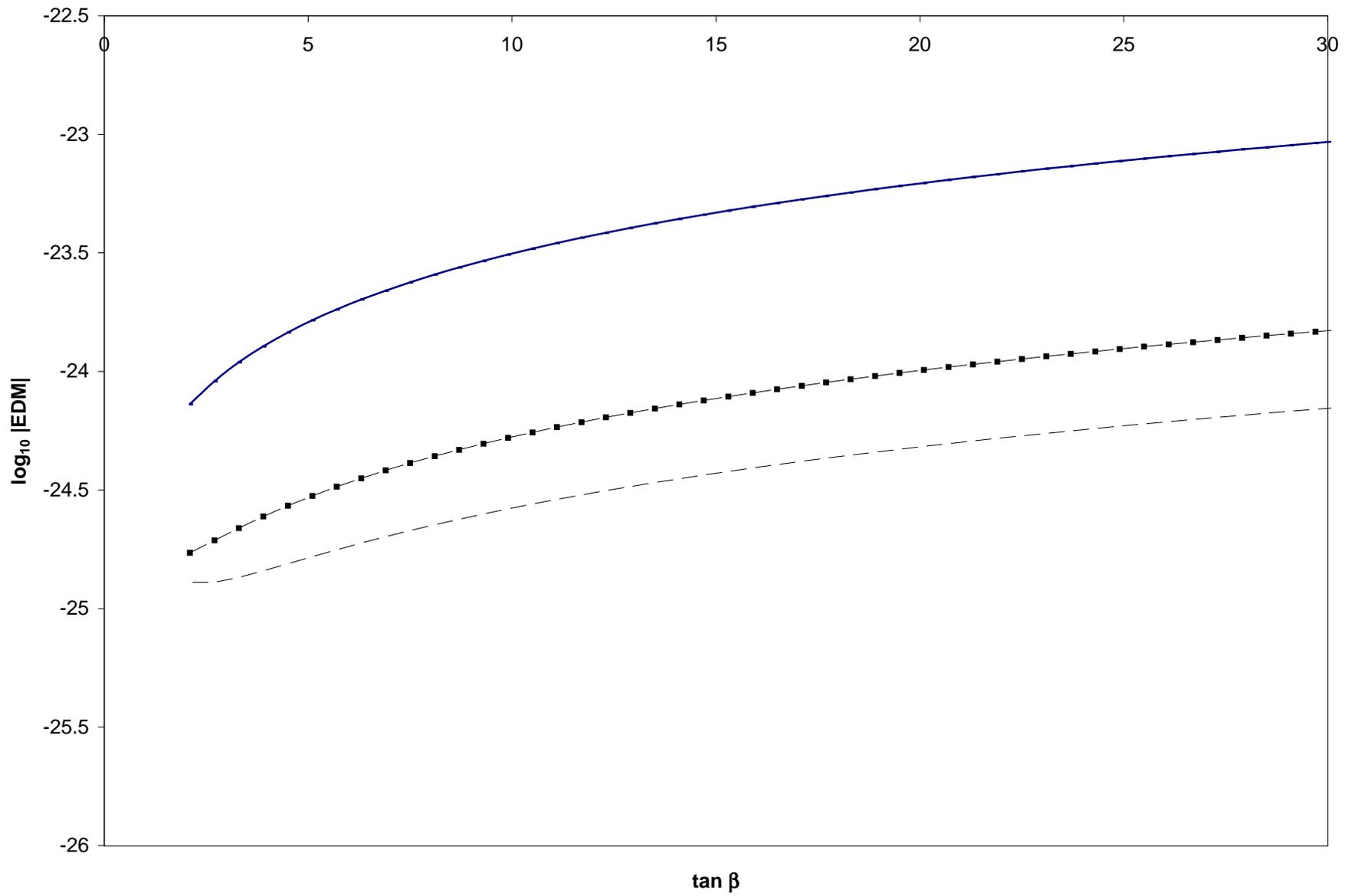
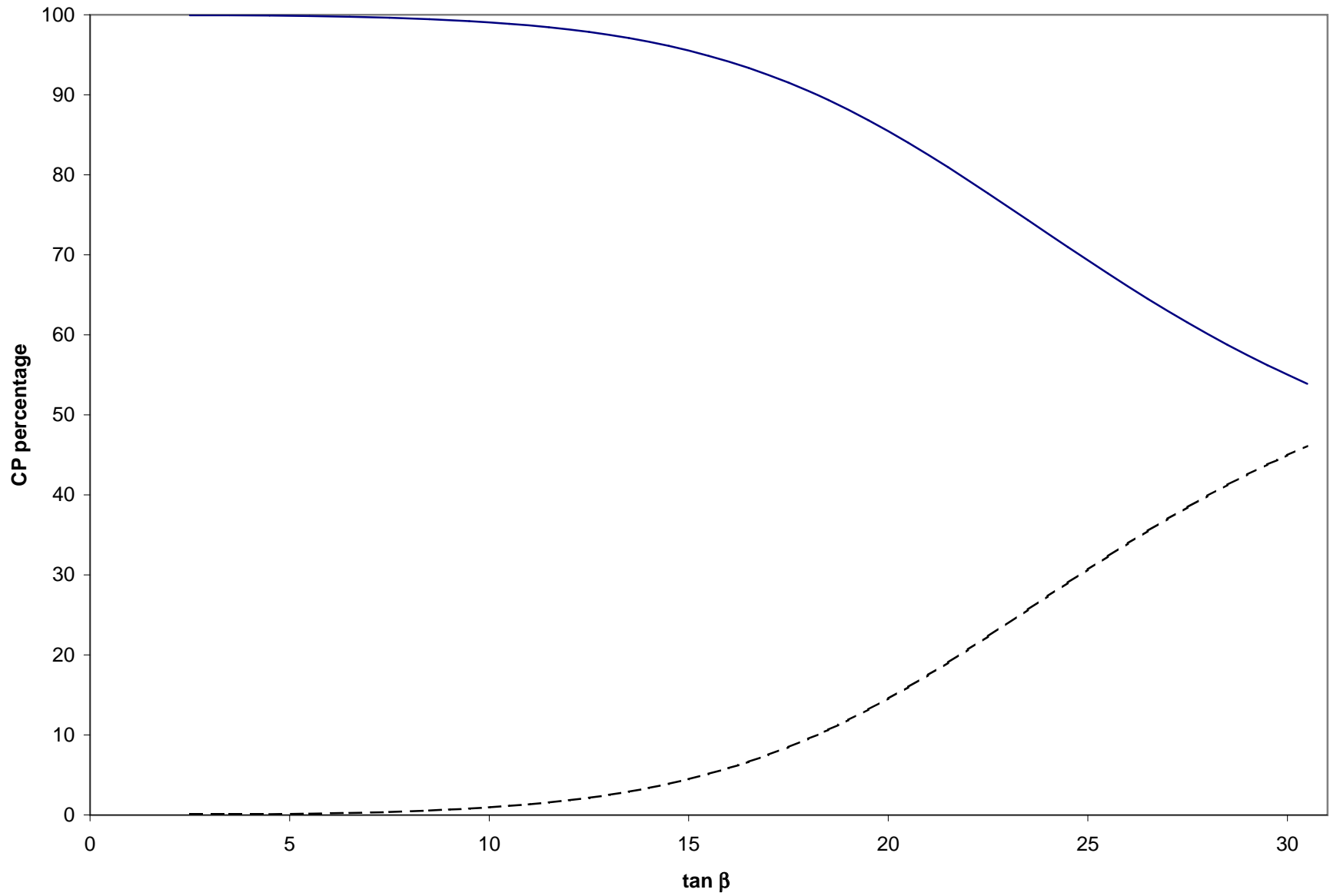
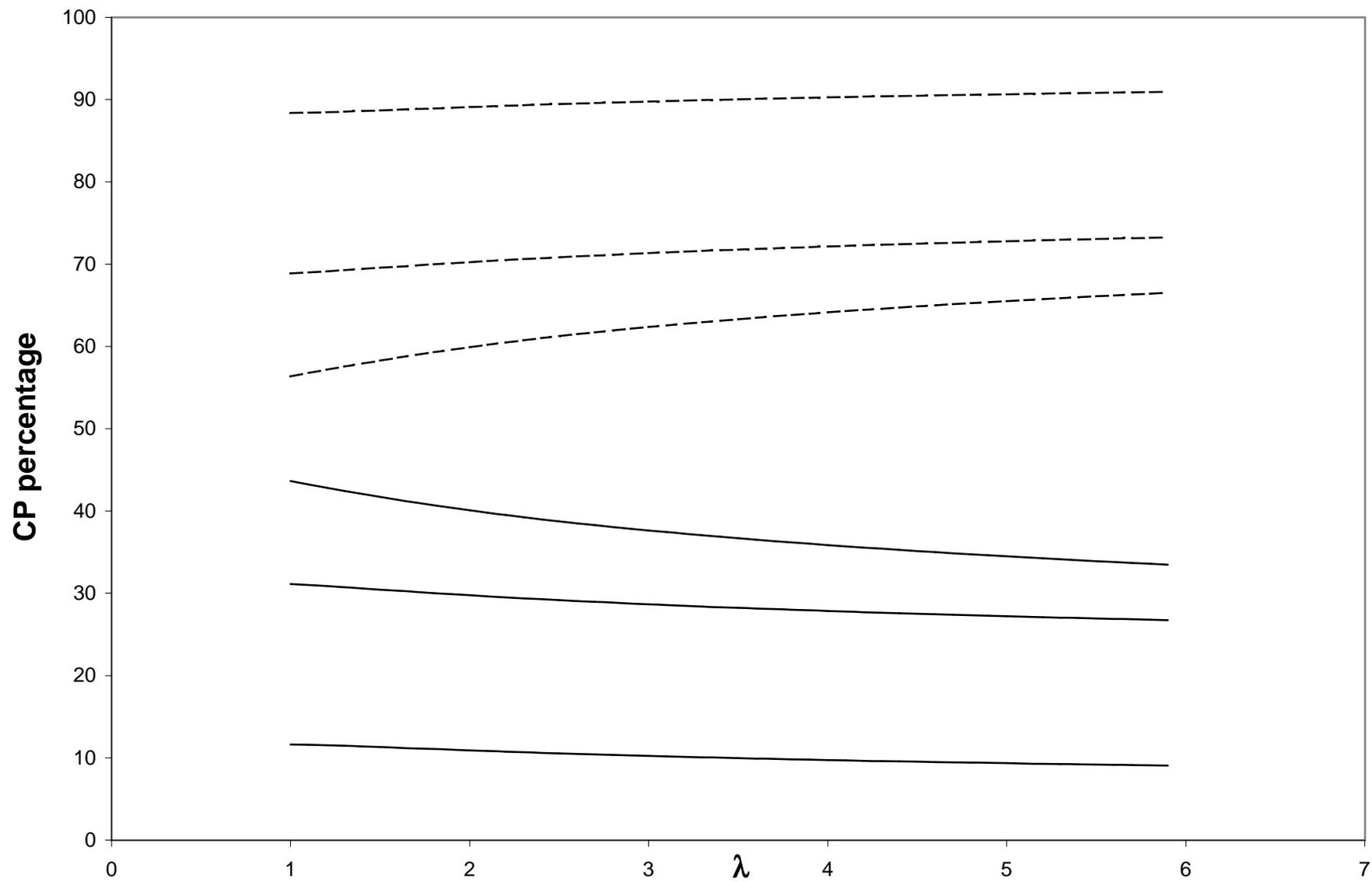
\end{document}